\documentclass[aps,prl,showpacs,twocolumn]{revtex4}
\usepackage{graphicx}
\begin{document}
\title{Unbiased Global Optimization of Lennard-Jones Clusters
 for $N \le 201$ by Conformational Space Annealing Method}
\author{Julian Lee$^{1,2}$, In-Ho Lee$^{1,3}$, and Jooyoung Lee$^1$\footnote{correspondence to: jlee@kias.re.kr}}
\affiliation{$^1$School of Computational Sciences, Korea Institute for Advanced Study,
Seoul 130-012, Korea   \\
$^2$ Department of Bioinformatics and Life Science, Soongsil University,
Seoul 156-743, Korea\\
$^3$ Korea Research Institute of Standards and Science,
  Daejon 305-600, Korea }
\begin{abstract}
  We apply the conformational space annealing (CSA) method
to the Lennard-Jones clusters
and  find all known lowest energy configurations up to 201 atoms,
{\it without using extra information of the problem} such as
the structures of the known global energy minima.
In addition, the robustness of the algorithm with respect to the
randomness of initial conditions of the problem is demonstrated by ten successful independent runs up to 183 atoms.
Our results indicate that the CSA method is a
general and yet efficient global optimization algorithm applicable to
many systems.
\end{abstract}
\maketitle
\setcounter{equation}{0}
  Finding the global minimum (GM) of a given function,
  called the global optimization, is an important problem in various
 fields of science and engineering. One of the simplest algorithms for global optimization is
the simulated annealing (SA) method \cite{SA}, which has been applied to many systems. Although SA
is very versatile in that it can be applied to many problems, the drawback is that its efficiency is usually much lower than
problem specific algorithms.
This is especially problematic for NP-hard problems such as protein folding or molecular cluster optimizations.
For this reason, it is important to find an algorithm
which is as general as SA,
and yet competitive with problem specific ones.

Recently, a powerful global optimization method called conformational space annealing (CSA) was
proposed \cite{CSA1},
and applied extensively and exclusively to the protein folding problem \cite{CSA2,CSA3,CSA4,CSA5,CSA6,CSA7,CSA8}.
The benchmark tests \cite{CSA1,CSA2,CSA3} have demonstrated
that it can not only find the known GM conformations
with less computations than existing algorithms,
but also provide new GMs in some cases \cite{CSA2,CSA3}.

The CSA unifies the essential ingredients of three global optimization methods,
Monte Carlo with minimization (MCM) \cite{MCM}, genetic algorithm (GA)~\cite{GA}, and SA.
First, as in MCM, we consider only the
phase space of local minima, i.e.~all configurations are energy-minimized by a
local minimizer.
Secondly, as in GA, we consider many configurations
(called {\it bank} in CSA) collectively,
and we perturb a subset of bank configurations ({\it seeds})
using other bank configurations.
This procedure is similar to mating in GA.
However, in contrast to the typical mating procedure in GA,
we often replace {\it small} portions of a seed
with the corresponding parts of bank configurations
in order to search the neighborhood of the seed configuration,
as will be elaborated later.
Finally, as in SA, we introduce a parameter $D_{\rm cut}$,
which plays the role of the temperature in SA.
The diversity of sampling is directly controlled in CSA
by introducing a distance measure between two configurations
and comparing it with $D_{\rm cut}$,
whereas in SA there are no such systematic controls.
The value of $D_{\rm cut}$ is slowly reduced just as in SA,
hence the name conformational space annealing.
Maintaining the diversity of the population using a distance measure was also tried in the context of GA~\cite{hart}, although no annealing was performed.

It should be noted that although the CSA method has been applied,
so far, only to the polypeptide systems,
the structure of the algorithm is not specific to these systems.
In fact, to apply the CSA to an optimization problem,
only two things are necessary;
a method for perturbing a seed configuration,
and a distance measure between two configurations.
This suggests that the CSA is a candidate for a versatile
and yet powerful global optimization method.
In this Letter, we demonstrate it by applying the CSA
to Lennard-Jones (LJ) clusters.

The LJ cluster is the system consisting
 of identical atoms interacting by a pairwise LJ potential:
\begin{equation}
 E(\{{\bf r } \}) = 4 \epsilon \sum_{i=1}^{N-1} \sum_{j=i+1}^N \big((\frac{\sigma}{r_{ij}})^{12} -(\frac{\sigma}{r_{ij}})^6\big),
\end{equation}
 where $N$ is the number of atoms and $r_{ij}$ the distance between atoms $i$ and $j$. We use the reduced unit where $\epsilon = \sigma =1$.
 It is not only interesting as a model for heavy inert gases,
 but also serves as a popular benchmark
system for optimization algorithms.
In fact, despite the simple form of the
interaction, finding the GM configuration has been a challenging problem
even for small $N$ \cite{north,free,farg,will,maranas,gomez,col,xue,pill,doye1,doye2,hart1,dh,leary1,barron,wales,wolf,white,leary2,scheraga,hart,barron2,rbg,leary3,krivov,hart3}.

Many of the powerful global optimization algorithms applied to this system are specific to the LJ cluster.
In particular,  they use the information on the structure
 of the known GMs of LJ clusters, favoring closely packed ones. For example, many GMs for
 $N \le 147$ were discovered for the first time in Ref.~\cite{north} by using icosahedrally derived lattices.
  A similar strategy \cite{rbg} was utilized for $N \le 309$.
Although these approaches are powerful, they fail when the GM configuration is
of an unexpected structure.
Also, since they are designed specifically for the LJ cluster problem, it is not clear if they can be applied to
 other systems.

 Recently there were a number of successful applications of unbiased search methods to LJ clusters.
 One of the most powerful
 methods in this category is the basin-hopping method \cite{wales,leary2}.
Almost all GM
 known at present \cite{url} were reproduced for $N \le 110$, and those for $N=69, 78, 107$ were updated. This is an impressive result indeed. However, there are certain magic
numbers ($N=76,77,103,104$) where the GMs could not be found by directly optimizing the system consisting of $N$ atoms. They could be found only by adding and subtracting an atom from the GM configurations of cluster size $N\pm1$. In Ref.~\cite{leary3}, a variant of the basin-hopping method was applied for $N \le 110$, which performed better than the original version, but the known lowest energy minima for $N=75-77$ could be found only $4,2,8$ times out of 1000 independent runs. Similarly, the unbiased search in Ref.~\cite{scheraga} could not reproduce the GMs for $N=75-77,98$. In Ref.~\cite{hart}, all GMs for $N \le 150$ were reproduced with a unbiased search method.
The same method was also applied for larger cluster sizes, and found some of the known lowest energy minima for $N \le 309$ \cite{hart3,url}.
A global optimization method which combines the idea of basin-hopping and the genetic algorithm was applied for the cluster size $N \le 309$ but failed to reproduce the known minima \cite{url} for $N=185$, $187$.

 In this work, we apply the CSA to the LJ cluster problem.
 We find all known
  GM configurations for $N \le 201$ \cite{url}.
  In particular, for each LJ cluster for
  $N \le 183$, we generate
  ten independent random configurations and succeed in finding the GMs for all cases  {\it without an exception}. This is an exhaustive test unprecedented in the literature, and shows that our algorithm is quite robust with respect to the change of initial conditions.


To elaborate on the details of the CSA method,  we first randomly generate a certain number of initial
configurations (50 in this work)  whose energy is subsequently
minimized. We call the set of these configurations the {\it first bank}. We make a copy of the
first bank and call it the {\it bank}. The configurations in the bank are updated in later stages, whereas those in the first bank are kept unchanged. Also, the number of configurations in the bank is kept unchanged when the bank is updated. The initial value of $D_{\rm cut}$ is set as $D_{\rm ave}/2$ where $D_{\rm ave}$ is the average distance between the configurations in the first bank.
New configurations are generated by choosing a certain number
(20 in this work) of {\it seed} configurations
from the bank and by replacing parts of the seeds
by the corresponding parts of configurations randomly chosen
from either the first bank or the bank.
Random perturbations are also performed.
In this work 20 and 10 configurations are generated
for each seed using the partial replacements and random perturbations,
respectively.
Then the energies of these configurations are subsequently minimized
(trial configurations).

A newly obtained local minimum configuration $\alpha$ is compared with those in the bank to decide how the bank
should be updated. One first finds the configuration $A$ in the bank which is closest to $\alpha$ with the distance $D(\alpha, A)$. If $D(\alpha, A) < D_{\rm cut}$, $\alpha$ is considered as similar to $A$. In this case, the configuration with lower energy among $\alpha$ and $A$ is kept in the bank, and the other one is discarded. However, if $D(\alpha, A) > D_{\rm cut}$, $\alpha$ is regarded as distinct from all configurations in the bank. In this case, the configuration with the highest energy among the bank configurations plus $\alpha$ is discarded,
and the rest are kept in the bank.
We perform this operation for all trial configurations.

This process of generating new conformations by perturbation and subsequent local minimizations, and updating the bank, can be visualized follows (Fig.1).
Each of the bank configurations can  be considered to represent
all local minima contained in
the sphere with radius $D_{\rm cut}$, centered on it.
To improve a bank configuration $A$,
we first select $A$ as a seed.
We perturb $A$ and subsequently energy-minimize it to generate a trial configuration $\alpha$.
When $\alpha$ originates from $A$ by small perturbation,
it is likely that $\alpha$ is
contained in a sphere centered on $A$.
If $\alpha$ replaces $A$, the center of
the sphere moves from $A$ to $\alpha$.
If $\alpha$ belongs to a different sphere centered on $B$,
$\alpha$ can replace $B$ in a similar manner.
When $\alpha$ is outside of all existing spheres,
a new sphere centered on $\alpha$ is generated.
In this case, to keep the number of
spheres fixed, we remove the sphere represented by the highest-energy
configuration.
Obviously, the former two cases are more likely to happen
when the spheres are large, and the latter when spheres are small.
Larger value of $D_{\rm cut}$ produces more diverse sampling,
whereas smaller value results in quicker search of low-energy configurations
at the expense of getting trapped in
a basin probably far away from the GM.

Therefore,
for efficient sampling of the phase space, it is necessary to
maintain the diversity of sampling in the early stages and then
gradually shift the emphasis toward obtaining low energy configurations,
by slowly reducing $D_{\rm cut}$.
 In practice, the $D_{\rm cut}$ is reduced by a fixed ratio after the bank update has been attempted by all the newly generated trial configurations,
 in such a way that $D_{\rm cut}$ reaches
 $D_{\rm ave}/5$ after $10000$ local minimizations.
 Then seeds are selected again from the bank configurations
 which have not been used as seeds yet, to repeat aforementioned procedure.
The value of $D_{\rm cut}$ is kept constant after it reaches the final value.

When the energy of a seed configuration does not improve
after a fixed number of perturbations,
we stop perturbing it.
To validate this judgment,
it is important that typical perturbations are kept small,
so that the perturbed configurations are close to their
original seeds. However, large perturbations are also performed, in order to efficiently sample various regions of the search space.

It should be noted that in the early stages of CSA the seed configurations are continuously being replaced by low energy local minima close to it.
Therefore, when all of the bank configurations are used as seeds (one iteration completed), usually after tens of thousands of local minimizations, this implies that the procedure of updating the bank might have
reached a deadlock. However, we give these configurations another chance by resetting them to be eligible for seeds, and repeat another iteration of search.
After a preset number of iterations,
we conclude that our procedure has reached a deadlock.
When this happens,
we enlarge the search space by adding more random configurations
into the bank and repeat the whole procedure
until the stopping criterion is met.
 In this work, after 3 iterations are completed,
 we increase the number of bank configurations by adding
 50 randomly generated and minimized configurations into the bank
 (and also into the first bank),
 and reset $D_{\rm cut}$ to $D_{\rm ave}/2$.
The algorithm stops when the known GM \cite{url} is found, which is examined after all the new trial configurations are used for possible bank updates.

It should be noted that since one iteration is completed
only after all bank configurations have been used as seeds,
and we add random configurations whenever our search has reached a deadlock,
there is no loss of generality for using particular values for
the number of seeds, the number of bank configurations, etc.

\begin{figure}
\includegraphics{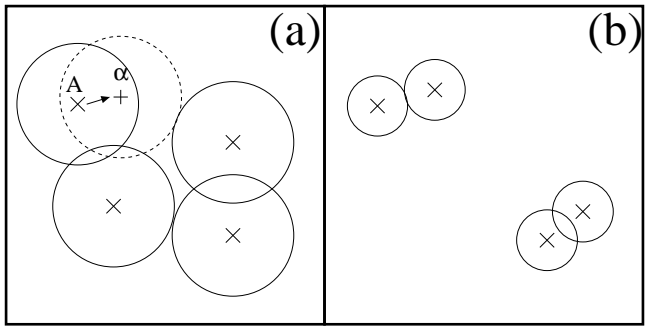}
\caption{\label{fig:rad} Schematic figure showing the search procedure of CSA. The boxes represent the identical phase space. (a) Initially, we cover the phase space by large spheres centered on randomly chosen local minima denoted by $\times$,
and replace the centers with lower-energy local minima. When $A$ is replaced by $\alpha$, the sphere moves in the direction of the arrow.
(b) As the algorithm proceeds and the energies of the representative configurations at the centers of the spheres are lowered, the size of the spheres are reduced and the search space is narrowed down to small basins of low-lying local minima.}
\end{figure}

We define the distance measure $D(k,k')$ as follows. Given
a configuration $k$, we define the first and second shell centered on
each atom as the spheres of radii 1.35 and 1.70.
These values of the radii are obtained from the radial distribution function,
which is approximately estimated from randomly generated
local minimum configurations with $N=201$, and do not depend much on $N$.
We then obtain the histogram
$H^k(1,n)$ ($H^k(2,n)$) which is the number of atoms having $n$
atoms in the first (second) shell.
The first and second coordination numbers
contain the geometrical information of the local
environment of each atom.
Therefore, the histograms provide collective information
of such local structures in a cluster.
We put larger weight for the first coordination number than the second one
since the former is more important than the latter.
In addition, larger weights are assigned to the histograms with larger
$n$, since they correspond to the core part of a cluster.
This motivates us to define $D(k,k')$ as:
\begin{figure}
\includegraphics{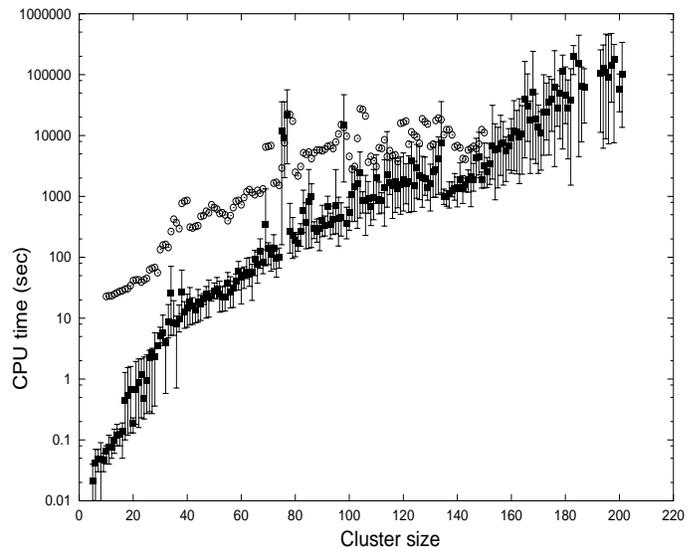}
\caption{\label{fig:cpu} CPU time (sec) of ten separate runs for
cluster size $N \le 201$.
The square boxes denote the average values and the error bars
indicate the ranges of the values.
The partially successful runs ($N=184,188-192,198,199$)
are excluded (See text).
The circles denote the result for $N \le 150$ reported in Ref.~\cite{hart} after rescaling of the CPU clocks.}
\end{figure}
\begin{eqnarray}
 D(k,k') &=&  \sum_n n \Big(2 | H^k(1,n)- H^{k'}(1,n) | \nonumber \\
 &&+ | H^k(2,n)- H^{k'}(2,n) | \Big) .
\end{eqnarray}
It should be noted that the details of the definition of
$D(k,k')$ do not affect the overall performance.

The method of perturbing a seed configuration $s$ to
generate a new configuration is as follows \cite{dh}. We first generate random planes passing
through the center of mass of the seed $s$
and another bank configuration $k$. We then choose from $s$ a certain
fraction ($25-50\%$) of atoms which are farthest from the plane, and replace
them by the corresponding counterpart in $k$. We
also make a random rotation perpendicular to the face of the cut
surface when we put together the two parts.
The main difference from the method used in Ref.~\cite{dh} is
that we are making a fractional replacement of a seed,
for the reason discussed earlier.
In addition,
we also generate new configurations
by choosing an atom in $s$ with the lowest
coordination number, and move it to the neighborhood of an atom with the second lowest coordination number.

We could reproduce all published GM
configurations \cite{url} for $N\le201$.
To measure the overall efficiency of our algorithm,
we performed exhaustive systematic runs.
We performed ten independent CSA runs for each cluster size $N \le 183$,
and found all known GMs for all ten cases, {\it without an exception},
which demonstrates the robustness of the algorithm
with respect to the randomness of the initial conditions.
The same test was performed for $184 \le N \le 201$,
but for $N=184,188-192,198,199$
the optimization was only partially successful
(4-9 out of 10).
The average values and fluctuations of CPU times for these runs
are shown in Fig.~\ref{fig:cpu}.
The computations were carried out on Athlon processors (1.667 GHz), and the limited-memory quasi-Newton method \cite{lbfg} was used for local minimization. The minimization stopped whenever $|G|/\sqrt{3N} \le 0.001$, where $G$ is the gradient.
A similar plot was presented in Ref.~\cite{hart} for $N \le 150$,
where the identical local minimizer was used.
These results are also included in Fig.~\ref{fig:cpu} for comparison,
after the rescaling of the CPU clocks.
The result suggests that our algorithm is faster on average
for the sizes where the data is available,
although a rigorous conclusion is difficult to be drawn due to the
possible technical differences such as the stopping criteria of the local
minimizer, and the fact that our result is the average of ten independent runs.

 The CSA method is now also being applied to quite different kinds of problems such as
the traveling salesman problems, spin glasses, not to mention complex molecular clusters. For example, the shortest path of ATT532 \cite{att}
was found for all 100 independent runs, whose results will be reported elsewhere.
This suggests that the CSA method is a
general and yet efficient global optimization algorithm applicable to
many systems. As is the case for GA,
the CSA can also be easily adapted for efficient parallel computation\cite{CSA3,CSA8}.
\bibliography{LJ}

\end{document}